\documentclass[conference]{IEEEconf}

\input epsf
\usepackage{graphicx}
\usepackage{multirow}
\usepackage{titlesec}
\usepackage{balance} 
\usepackage{stfloats}
\usepackage{xcolor}
\usepackage{algorithm}
\usepackage{algorithmicx}
\usepackage{algpseudocode}
\usepackage{amsmath}

\renewcommand\thesection{\arabic{section}}
\renewcommand\thesubsectiondis{\thesection.\arabic{subsection}}
\renewcommand\thesubsubsectiondis{\thesubsectiondis.\arabic{subsubsection}}
\begin{document}

\title{\textbf{\Large Hybrid Privacy Policy-Code Consistency Check using Knowledge Graphs and LLMs\\}}

\author{Zhenyu Mao$^{1}$, Xinxin Fan$^{1}$, Yifei Wang$^{1}$, Jacky Keung$^{1,*}$, and Jialong Li$^{2}$\\
	\normalsize $^{1}$City University of Hong Kong, Hong Kong, China\\
	\normalsize $^{2}$Waseda University, Tokyo, Japan\\
	\normalsize zhenyumao2-c@my.cityu.edu.hk, xinxinfan3-c@my.cityu.edu.hk\\
        \normalsize ywang4748-c@my.cityu.edu.hk, lijialong@fuji.waseda.jp\\
	\normalsize *corresponding author: Jacky.Keung@cityu.edu.hk
}

\maketitle
\begin{abstract}

The increasing concern in user privacy misuse has accelerated research into checking consistencies between smartphone apps' declared privacy policies and their actual behaviors.
Recent advances in Large Language Models (LLMs) have introduced promising techniques for semantic comparison, but these methods often suffer from low accuracies and expensive computational costs.
To address this problem, this paper proposes a novel hybrid approach that integrates 1) knowledge graph-based deterministic checking to ensure higher accuracy, and 2) LLMs exclusively used for preliminary semantic analysis to save computational costs.
Preliminary evaluation indicates this hybrid approach not only achieves 37.63\% increase in precision and 23.13\% increase F1-score but also consumes 93.5\% less tokens and 87.3\% shorter time.

\end{abstract}
\IEEEoverridecommandlockouts
\vspace{1.5ex}

\IEEEpeerreviewmaketitle

\section{Introduction}
\label{sec:1}

Smartphone apps have transformed modern life, influencing nearly every aspect of human activity ranging from daily purchases to civic engagement.
These apps typically justify their extensive data collection practices as necessary for service improvement, however, recently there is a rising concern in the misuse of these data.
While developers are required to disclose their data practices through privacy policies, numerous studies have highlighted significant gaps between these formal declarations and the apps' actual data handling behaviors \cite{javed2024systematic}.

Privacy policy-code consistency check has emerged as a significant research stream \cite{tan2023ptpdroid, zhao2023demystifying}, where most studies employ static analysis tools like FlowDroid for this task \cite{javed2024systematic, arzt2014flowdroid}.
Recent work has explored LLM-based approaches for semantic comparison \cite{morales2024large}, typically implementing a pipeline that: 1) extracts events and data flows, 2) identifies relevant methods, and 3) computes policy-code similarity.
However, the exclusive LLMs-based approach suffers from two fundamental limitations: 1) Over-alignment bias in LLMs frequently generates false positives by fabricating policy-code relationships, reducing checking accuracy. 2) End-to-end LLMs-based processing requires iterative prompting sequences, resulting in higher computational costs.

To address this problem, this paper proposes a hybrid approach that combines: 1) knowledge graph-based deterministic checking to avoid false positives caused by over-alignment bias, and 2) LLMs focusing exclusively on preliminary semantic analysis prior to the final comparison, reducing the prompting.

\section{Proposed Hybrid Approach}
\begin{figure}[hbtp]
    \centering
    \includegraphics[height=50mm]{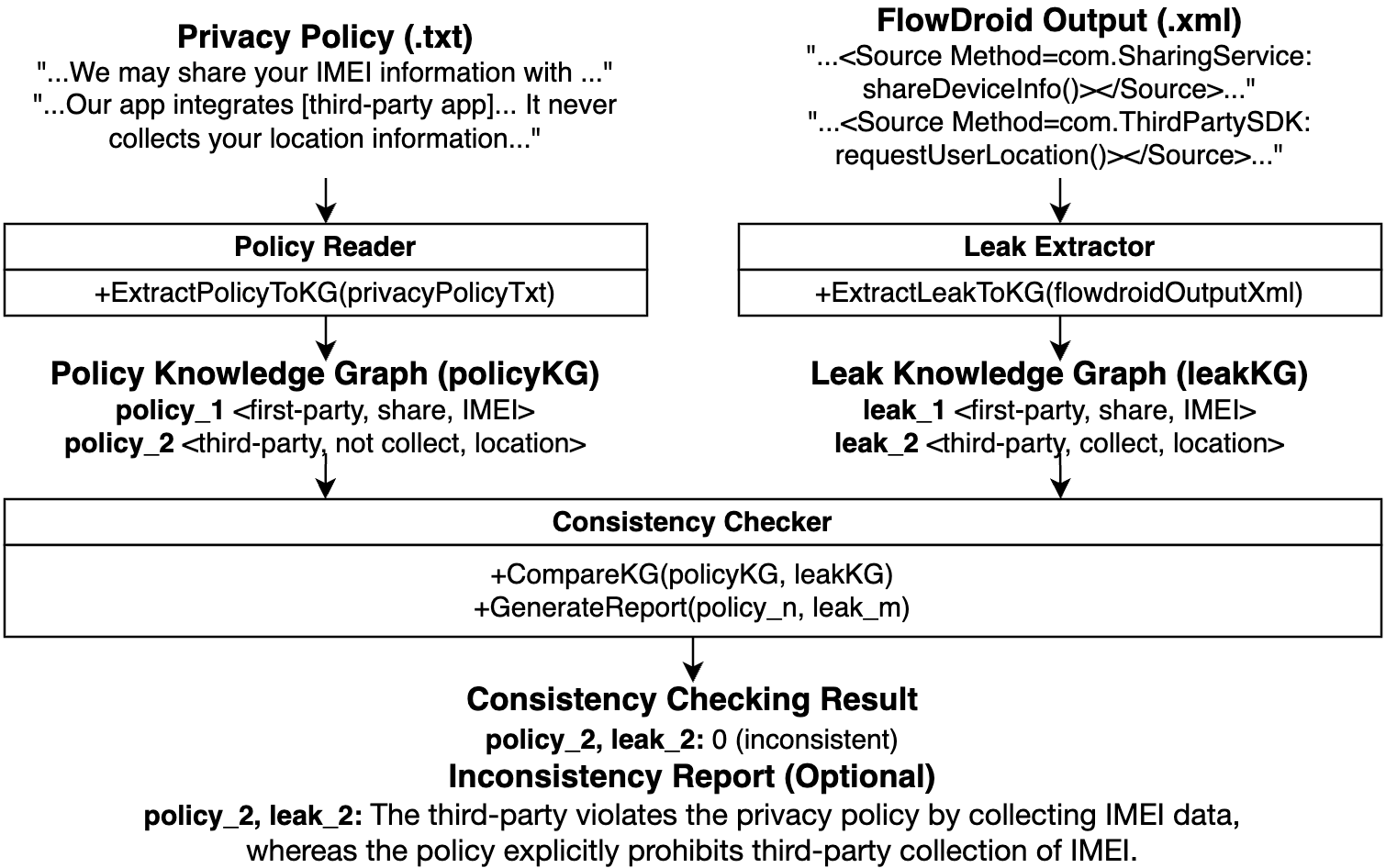}
    \caption{Overview of Hybrid Approach}
    \label{fig:1}
\end{figure}

Figure \ref{fig:1} presents an overview of the hybrid approach, realized through coordinated interaction among three core components: a Policy Reader, a Leak Extractor, and a Consistency Checker.

\textbf{Policy Reader:}
The policy reader employs an LLM to transform the natural language-based privacy policies into a structured knowledge graph (i.e., $policyKG$).
Each $policyKG$ triple follows the format $<actor,action,data>$, where: $actor$ denotes the responsible entity as either first-party (i.e., the app) or third-party (e.g., SDKs), $action$ represents data operations, classified as $collect$, $share$, or their negation via LLM semantic mapping, and $data$ indicates the processed user data type, standardized through LLM mapping according to a list containing $14$ common data types, referring to \cite{tan2023ptpdroid}.

\textbf{Leak Extractor:}
The leak extractor extracts methods containing potential violations by processing FlowDroid's XML output, which encodes static analysis results of sensitive data flows in android apps.
Through LLM semantic analysis, it constructs a leak knowledge graph (i.e., $leakKG$) that mirrors the $policyKG$ structure, enabling direct consistency checking.

\textbf{Consistency Checker:}
The consistency checker executes the checking algorithm shown in Algorithm \ref{algo:1} to detect inconsistency between $policyKG$ and $leakKG$ triples.
For each identified inconsistency (i.e., flagged as $False$), the consistency checker optionally invokes an LLM to generate a natural language inconsistency report, providing a detailed explanation of how the app's actual behavior violates its privacy policy.

\begin{algorithm}
\caption{Consistency Checking Algorithm}
\begin{algorithmic}[1]
\State \textbf{Input:} $policyKG,\ leakKG$
\For{each $l \in leakKG$}
    \For{each $p \in policyKG$}
        \If{$p.actor == l.actor \wedge p.data == l.data$}
            \If{$p.action == l.action$}
                \State \Return $True,\ l,\ p$
            \ElsIf{$p.action==(!l.action)$}
                \State \Return $False,\ l,\ p$
            \EndIf
        \EndIf
    \EndFor
    \State \Return $False,\ l,\ Null$
\EndFor
\end{algorithmic}
\label{algo:1}
\end{algorithm}

\section{Preliminary Evaluation}
This preliminary evaluation is driven by the following two Research Questions (RQs):
\textbf{RQ1:} Compared to the pure LLMs approach, does the hybrid approach provide higher accuracy?
\textbf{RQ2:} Compared to the pure LLMs approach, does the hybrid approach reduce computational costs in time and tokens?

\subsection{Experiment Settings}
\textbf{Raw data:}
The experiment uses FlowDroid-analyzed outputs from 7 Android apps with their privacy policies, supplemented by author-constructed test cases (totaling 17 XML files and 23 policies) to ensure full coverage of common data types in \cite{tan2023ptpdroid}.

\textbf{Ground Truth:}
The ground truth was established through dual-author annotation with consensus validation.

\textbf{Baseline:}
The proposed hybrid approach is compared against the pure LLM baseline as introduced in Section \ref{sec:1} \cite{morales2024large}.
LLMs used for this experiment are instances of DeepSeek-V3 \cite{deepseek2024}.

\textbf{Evaluation Metrics:}
To ensure a systematical evaluation, precision rate, recall rate, and F1 score, are employed to assess the accuracy in RQ1, while end-to-end execution time and token usage are used to quantify computational costs in RQ2.

\subsection{Experiment Results and Discussion}

\begin{table}[htbp]
\centering
    \caption{Comparison in precision, recall, and F1}
    \begin{tabular}{l|cc} \hline
        Metrics & Baseline & Hybrid \\\hline
        Precision (\%) & 42.37 & \textbf{80.00} \\
        Recall (\%) & \textbf{80.65} & 77.42 \\
        F1 (\%) & 55.56 & \textbf{78.69} \\\hline
    \end{tabular}
    \label{table:1}
\end{table}

\begin{table}[htbp]
\centering
    \caption{Comparison in total token and time consumption}
    \begin{tabular}{c|ccc} \hline
        Cost & Baseline & Hybrid w/o & Hybrid w/ \\
        types & approach & explanation & explanation \\\hline
        Prompt token & 381134 & \textbf{23668} & 60650 \\
        Completion token & 2674 & \textbf{1432}& 33888 \\
        Total token & 383808 & \textbf{25100} & 94538 \\\hline
        Time (sec) & 1625 & \textbf{207} & 2551 \\\hline
    \end{tabular}
    \label{table:2}
\end{table}

\begin{figure}[hbtp]
    \centering
    \includegraphics[height=12mm]{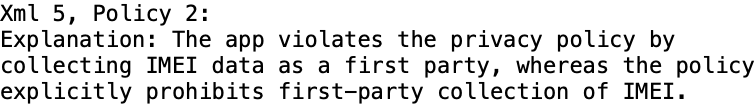}
    \caption{Example of Generated Inconsistency Report}
    \label{fig:2}
\end{figure}

As demonstrated in Table \ref{table:1}, the hybrid approach achieves significant improvements over the baseline, showing 37.63\% higher precision and 23.13\% better F1-score while maintaining competitive recall (77.42\%).
In terms of computational costs, as shown in Table \ref{table:2}, the hybrid approach outperformed with 93.5\% reduction in token usage and 87.3\% faster processing time from input to binary output.
The advantage in token usage even remains consistent with taking the tokens for generating inconsistency reports (as exemplified in Figure \ref{fig:2}) into account.

The experimental results demonstrate that the proposed hybrid approach successfully addresses two fundamental limitations in the pure LLM approach.
First, replacing LLMs-based semantic similarity matching with the knowledge graph-based deterministic checking managed to eliminate numerous false positives without significantly compromising recall rate by avoiding over-alignment bias.
Second, by restricting LLMs to preliminary semantic analysis prior to final comparison, computational costs in terms of both time and tokens are significantly reduced through minimized interactions with LLMs.

\section{Conclusion and Future Works}

This paper presents a novel hybrid approach that addresses key limitations of low accuracies and high computational costs in LLMs-based privacy policy-code consistency checking by integrating knowledge graph-based deterministic checking and LLMs exclusively for preliminary semantic analysis.
Preliminary results demonstrate significant improvements in both precision (increased by 37.63\%) and F1-score (increased by 23.13\%) while reducing computational costs (93.5\% fewer tokens and 87.3\% shorter time).
Future work should include: 1) regulatory compliance assessment against legal standards like GDPR, and 2) actionable repairing guidance for detected violations, enabling fully privacy policy-code alignment.

\bibliographystyle{IEEEtran}
\bibliography{QRS2025}

\begin{thebibliography}{1}
\providecommand{\url}[1]{#1}
\csname url@samestyle\endcsname
\providecommand{\newblock}{\relax}
\providecommand{\bibinfo}[2]{#2}
\providecommand{\BIBentrySTDinterwordspacing}{\spaceskip=0pt\relax}
\providecommand{\BIBentryALTinterwordstretchfactor}{4}
\providecommand{\BIBentryALTinterwordspacing}{\spaceskip=\fontdimen2\font plus
\BIBentryALTinterwordstretchfactor\fontdimen3\font minus \fontdimen4\font\relax}
\providecommand{\BIBforeignlanguage}[2]{{%
\expandafter\ifx\csname l@#1\endcsname\relax
\typeout{** WARNING: IEEEtran.bst: No hyphenation pattern has been}%
\typeout{** loaded for the language `#1'. Using the pattern for}%
\typeout{** the default language instead.}%
\else
\language=\csname l@#1\endcsname
\fi
#2}}
\providecommand{\BIBdecl}{\relax}
\BIBdecl

\bibitem{javed2024systematic}
Y.~Javed \emph{et~al.}, ``A systematic review of privacy policy literature,'' \emph{ACM Computing Surveys}, 2024.

\bibitem{tan2023ptpdroid}
Z.~Tan \emph{et~al.}, ``Ptpdroid: Detecting violated user privacy disclosures to third-parties of android apps,'' in \emph{ICSE}, 2023.

\bibitem{zhao2023demystifying}
K.~Zhao \emph{et~al.}, ``Demystifying privacy policy of third-party libraries in mobile apps,'' in \emph{ICSE}, 2023.

\bibitem{arzt2014flowdroid}
S.~Arzt \emph{et~al.}, ``Flowdroid: Precise context, flow, field, object-sensitive and lifecycle-aware taint analysis for android apps,'' \emph{ACM sigplan notices}, 2014.

\bibitem{morales2024large}
G.~Morales \emph{et~al.}, ``A large language model approach to code and privacy policy alignment,'' in \emph{2024 SANER}, 2024.

\bibitem{deepseek2024}
DeepSeek, ``Deepseek-v3: Advanced language model api,'' \url{https://platform.deepseek.com}, 2024.

\end{thebibliography}

\balance

\end{document}